\documentclass[journal=jacsat,manuscript=article]{achemso}
\setkeys{acs}{etalmode=truncate,maxauthors=10,articletitle=true}

\usepackage[version=3]{mhchem} 
\SectionNumbersOn
\usepackage{textcomp, gensymb}
\usepackage{bm}
\usepackage{booktabs}
\usepackage{chemfig}
\usepackage{amsmath}

\author{Bowen Han}
\email{hanb1@ornl.gov}
\affiliation{Neutron Scattering Division, Oak Ridge National Laboratory, Oak Ridge, Tennessee 37830, USA}

\author{Yongqiang Cheng}
\email{chengy@ornl.gov}
\affiliation{Neutron Scattering Division, Oak Ridge National Laboratory, Oak Ridge, Tennessee 37830, USA}

\title[An \textsf{achemso} demo]
  {Benchmarking Universal Machine Learning Interatomic Potentials for Real-Time Analysis of Inelastic Neutron Scattering Data}

\begin{document}

\vfill
This manuscript has been authored by UT-Battelle, LLC under Contract No. DE-AC05-00OR22725 with the U.S. Department of Energy.  The United States Government retains and the publisher, by accepting the article for publication, acknowledges that the United States Government retains a non-exclusive, paid-up, irrevocable, world-wide license to publish or reproduce the published form of this manuscript, or allow others to do so, for United States Government purposes.  The Department of Energy will provide public access to these results of federally sponsored research in accordance with the DOE Public Access Plan (http://energy.gov/downloads/doe-public-access-plan).
\clearpage




\begin{abstract}

The accurate calculation of phonons and vibrational spectra remains a significant challenge, requiring highly precise evaluations of interatomic forces. Traditional methods based on the quantum description of the electronic structure, while widely used, are computationally expensive and demand substantial expertise. Emerging universal machine learning interatomic potentials (uMLIPs) offer a transformative alternative by employing pre-trained neural network surrogates to predict interatomic forces directly from atomic coordinates. This approach dramatically reduces computation time and minimizes the need for technical knowledge. In this paper, we produce a phonon database comprising nearly 5,000 inorganic crystals to benchmark the performance of several leading uMLIPs. We further assess these models in real-world applications by using them to analyze experimental inelastic neutron scattering data collected on a variety of materials. Through detailed comparisons, we identify the strengths and limitations of these uMLIPs, providing insights into their accuracy and suitability for fast calculations of phonons and related properties, as well as for real-time interpretation of neutron scattering spectra. Our findings highlight how the rapid advancement of AI in science is revolutionizing experimental research and data analysis.

\end{abstract}

\clearpage

\section{Introduction}
\label{sec:intro}

Atomistic modeling is an indispensable tool to connect theory and experiment and to illustrate fundamental mechanisms behind the structure-property relationships in materials. An essential step in atomistic modeling is the evaluation of interatomic interactions, which can be derived from the potential energy as a function of the atomic coordinates\cite{reissland_physics_1973}. Such calculations can be performed at different levels of accuracy and with different efficiency. A simple model describes interatomic interactions in a parameterized formula, e.g., by assuming that the atoms are connected by springs of specific elastic constants. These empirical models are quick to compute, but cannot describe complex many-body interactions. The potential energy and forces can also be calculated from density functional theory (DFT), by solving the electron density and wave functions of the ground state, as well as their dependence on the atomic displacements\cite{togo_first_2015}. These calculations based on quantum mechanics can be very accurate, but they are also computationally expensive. The accuracy-efficiency tradeoff leads to many practical challenges. For example, the analysis of vibrational or phonon spectroscopy, such as those measured from inelastic neutron scattering (INS)\cite{fultz_inelastic_2020}, requires high-precision calculations of interatomic interactions, which is usually nontrivial, time-consuming, and needs special computing resources and expertise. The analysis usually cannot be performed in real-time or on-demand, which is not only a bottleneck for productivity but also a barrier for future autonomous experiments, where understanding the results on-the-fly is a prerequisite. 

The recent development of neural networks and machine learning for materials science has led to a possible solution that enables a new approach to the calculation of interatomic forces: training a neural network surrogate to predict energies and forces of a given atomic structure using available DFT data\cite{unke_machine_2021}. These neural network surrogates are usually called machine learning force fields or machine learning interatomic potentials (MLIPs). Neural networks excel at capturing complex, high-dimensional, nonlinear mappings, making it possible for MLIPs to achieve both accuracy and efficiency. Although early MLIPs were developed for specific compositions or materials, the availability of larger databases with more comprehensive coverage, combined with the rapid growth of computing power, makes it possible to train foundation models, or universal MLIPs (uMLIPs), that can describe a wide range of materials. The first uMLIPs are benchmarked for energies and forces, as well as some other global properties such as modulus\cite{riebesell_matbench_2024}. However, studies found that accurate descriptions of phonons could be more challenging, and most early models systematically underestimate the frequencies\cite{deng_systematic_2025}. Several models published very recently seem to be better and could be ready for phonon calculations according to Loew et al.\cite{loew_universal_2024} Can we now use uMLIPs for the analysis of phonons and interpretation of vibrational spectral data?

In this work, we produce a new phonon database using DFT. The database covers nearly 5000 crystals with unit cells containing up to 12 atoms. Compared to existing phonon databases (such as the one by Togo\cite{togo_atztogophonondb_2025}), we focus on a more comprehensive coverage of crystals with relatively simple crystal structures, large Brillouin zones, and rich phonon dispersion features that reflect the subtle interatomic interactions. We then benchmark the best-performing uMLIPs against our DFT database. Our application scenario focuses on the analysis of INS data, for which modeling of phonons has been the primary bottleneck. We compare phonons and simulated INS spectra obtained from the uMLIPs with available experimental data. Finally, we include some of the latest uMLIPs in the INSPIRED software\cite{han_inspired_2024}, which has a simple graph user interface to allow easy and quick simulations of INS spectra from a structure model. The results of this work illustrate an approach to the real-time analysis and interpretation of INS data.

\section{Results and Discussion}
\label{sec:results}

The crystals included in the benchmarking database are chosen from the Materials Project\cite{jain_commentary_2013}. Only stable and realistic materials are considered. We build our database starting from smaller unit cells (ranked by the number of atoms in the unit cell, in ascending order, up to 12 atoms). Details on the production of the database can be found in Methods. The final database contains 4869 crystals that cover 86 elements.

We include 12 uMLIPs from the top of the MatBench Discovery leaderboard by April 2025\cite{riebesell_matbench_2024}. They are eSEN-30M-OAM\cite{fu_learning_2025}, ORB v3\cite{rhodes2025orbv3atomisticsimulationscale}, SevenNet-MF-ompa\cite{park_scalable_2024,kim_sevennet_mf_2024}, GRACE-2L-OAM\cite{bochkarev_graph_2024}, MatterSim v1 5M\cite{yang_mattersim_2024}, MACE-MPA-0\cite{batatia_foundation_2024, Batatia2022mace, Batatia2022Design}, eqV2 M\cite{barroso-luque_open_2024}, ORB v1\cite{neumann2024orbfastscalableneural}, SevenNet-0\cite{park_scalable_2024}, MACE-MP-0\cite{batatia_foundation_2024, Batatia2022mace, Batatia2022Design}, CHGNet\cite{deng_chgnet_2023}, and M3GNet\cite{chen_universal_2022}. For each crystal, with each uMLIP, we first relax the atomic coordinates to find the optimized structure with the minimum potential energy. This structure is compared with the DFT-relaxed structure, and differences in atomic coordinates are averaged to account for discrepancies in the ground-state structure (Figure 1a). Phonon calculations are then performed on a uniform mesh sampling of the first Brillouin zone, and the average differences in the phonon frequencies are compared in Figure 1b. Finally, the phonon density of states (PDOS) is compared for spectral similarity, quantified by Spearman coefficients\cite{baumann_computer-assisted_1997,zou_deep_2023} (Figure 1c). The Spearman coefficient ranges between 0 and 1, with higher values indicating better alignment between the two PDOS compared. We also calculate the vibrational entropy ($S$), Helmholtz free energy ($F$), and constant volume heat capacity ($C_V)$ from the phonon results, and they are compared in the Supplementary Information (Figure S1). Detailed statistics of the comparison among all uMLIPs are shown in Tables S1-S6. These benchmarking results suggest that the most accurate uMLIPs for phonon calculations are ORB v3, SevenNet-MP-ompa and GRACE-2L-OAM, with MatterSim 5M, MACE-MPA-0, and eSEN-30M-OAM following closely. Our benchmark also suggests that the latest uMLIPs outperform the top ones in an earlier benchmark performed by Loew et al.\cite{loew_universal_2024}, when these potentials had not been published.

\begin{figure}[h!tbp]
   \centering
    \includegraphics[width=0.95\textwidth]{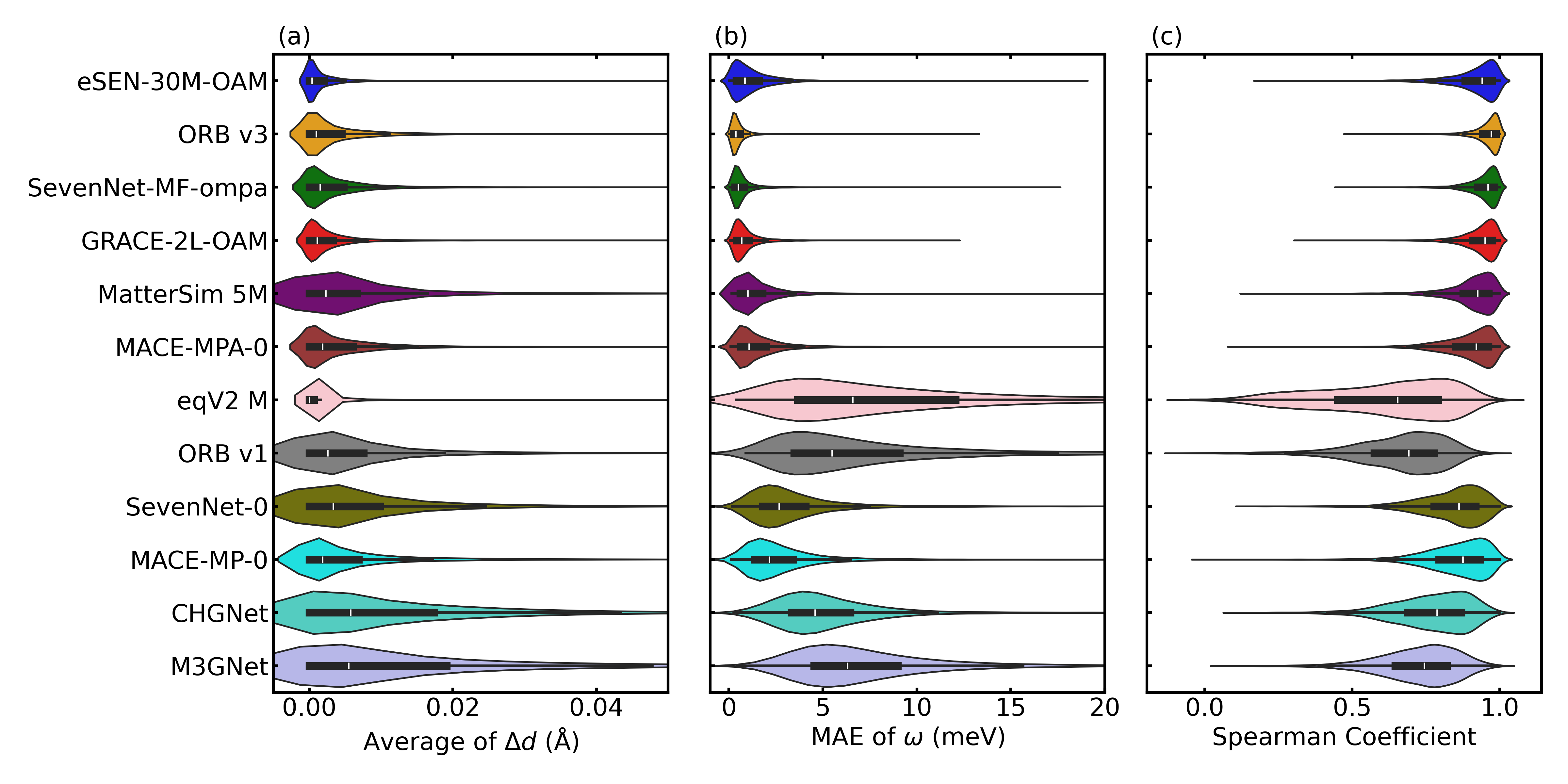}
    \caption{Violin plots of (a) differences in optimized atomic coordinates, (b) differences in phonon frequencies, and (c) Spearman coefficients of PDOS for the twelve uMLIPs benchmarked against DFT.}
    \label{fig:violin}
\end{figure}

The above analysis shows statistically how well each uMLIP performs against DFT. However, the ultimate benchmark should be the comparison with the experimental data. For example, in a real application scenario, we would compare the uMLIP calculations with the collected experimental data and use the model to provide quick analysis and interpretation by matching the simulated spectral features with the experimentally measured ones. The validation of the phonon calculation results with experimental data is also crucial for the prediction of phonon-based thermodynamic and kinetic properties. In the following, we will compare the performance of the uMLIPs using INS data as benchmarks. 

INS measures the dynamical structure factor $S(\bm{Q},E)$, which is a function of momentum $\bm{Q}$ (as a vector in the three-dimensional reciprocal space) and energy $E$. With orientational averaging for powder samples, the 4-dimensional (4D) $S(\bm{Q},E)$ reduces to 2D $S(\left|Q\right|,E)$, which is a function of the magnitude of $\bm{Q}$ (or $\left|Q\right|$) and energy $E$\cite{squires_introduction_2012}. Neutron scattering from a material can be coherent or incoherent. Some elements have intrinsically high incoherent neutron scattering cross-sections; thus materials containing such elements will have INS intensities mostly from the incoherent scattering. Some elements are strong coherent scatterers; thus scattering from these elements interferes with each other, exhibiting significant $\bm{Q}$-dependent patterns in the $S(\bm{Q},E)$ data. Note that when a crystal structure is complex with a large unit cell, even if the constituent elements are coherent scatters, the Brillouin zone may be so small that the $\bm{Q}$ dependence cannot be resolved after powder averaging. As a result, the overall scattering may appear to be incoherent. This is the so-called incoherent approximation\cite{squires_introduction_2012}. In the following, we will show examples for each scenario.

We start with a simple crystal of a coherent neutron scatterer: graphite. As a reference, the phonon dispersion curves calculated for graphite using various uMLIPs are compared with DFT in Figure 2a-f. A powder sample of graphite was measured at SEQUOIA\cite{granroth_sequoia_2010}, a direct geometry spectrometer at the Spallation Neutron Source, to obtain $S(\left|Q\right|,E)$\cite{cheng_simulation_2019}. Strong effects of coherent inelastic scattering can be seen even in the powder data, where various phonon dispersion curves can be resolved (Figure 2g). Compared with the experimental and DFT-calculated INS spectra, it is clear that ORB v3 and MatterSim are among the best to capture the details in the powder $S(\left|Q\right|,E)$. However, ORB v3 produces ZA branches with slight negative frequencies, whereas MatterSim predicts stable ZA modes. Negative frequency modes are also observed with SevenNet and GRACE, but not for the two MACE potentials. The negative frequencies indicate phonon instability and can be detrimental to the simulated INS spectra at low energy. We also note that the phonon frequencies calculated by MACE-MP-0 are significantly underestimated. The observations in the simulated INS spectra are consistent with the comparisons of calculated phonon dispersion.

\begin{figure}[h!tbp]
   \centering
    \includegraphics[width=0.95\textwidth]{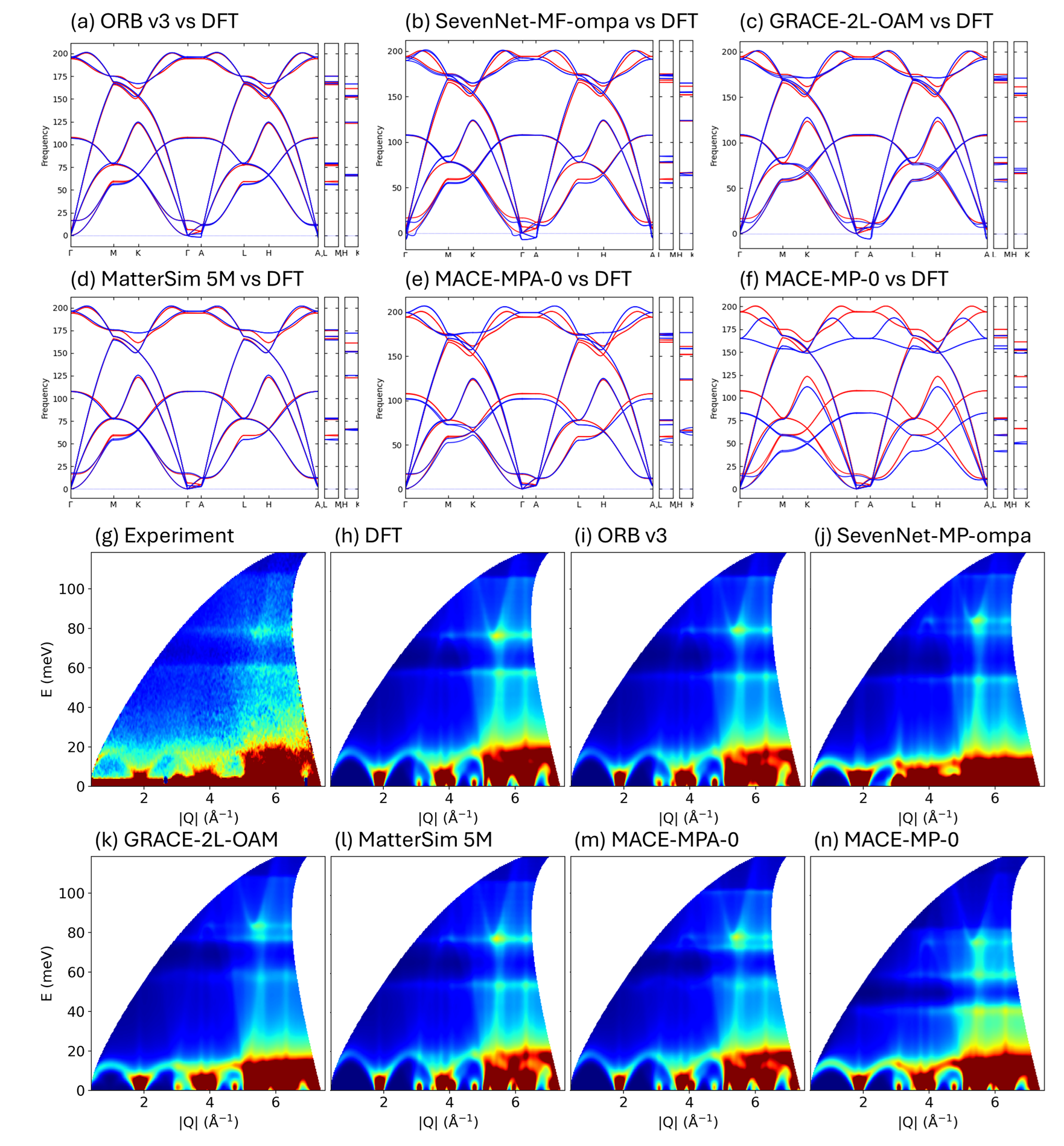}
    \caption{(a)-(f) Phonon dispersion of graphite obtained from uMLIPs and DFT. The powder INS spectra measured at SEQUAIA is shown in (g)\cite{cheng_simulation_2019}, with the corresponding simulations in (h)-(n).}
    \label{fig:graphite}
\end{figure}

Next, we move to a more complex crystal, \ce{Ba3ZnRu2O9},\cite{cheng_direct_2023} that has 30 atoms in the unit cell. Phonon dispersion plots in Figure 3a-f show more complex features and obvious discrepancies between uMLIP and DFT in all models. Specifically, the main issue with most uMLIPs is still the systematic softening of the calculated frequencies in the high-frequency optical band. Two models, ORB v3 and MatterSim, outperform others in this regard, reproducing the highest-frequency branches with better agreement. Comparison of powder INS spectra in Figure 3g-n reveals the same trend. The more complex crystal structure results in weaker coherent scattering patterns in the powder $S(\left|Q\right|,E)$, preventing a direct comparison of specific phonon dispersion curves. However, the energy levels of multiple phonon bands are clearly visible and can be used to evaluate the accuracy of different models. Note that the feature at below 1 \AA$^{-1}$ in the experimental spectrum is due to magnetic scattering, which is not included in the simulated ones where only intensities from phonons are calculated.

\begin{figure}[h!tbp]
   \centering
    \includegraphics[width=0.95\textwidth]{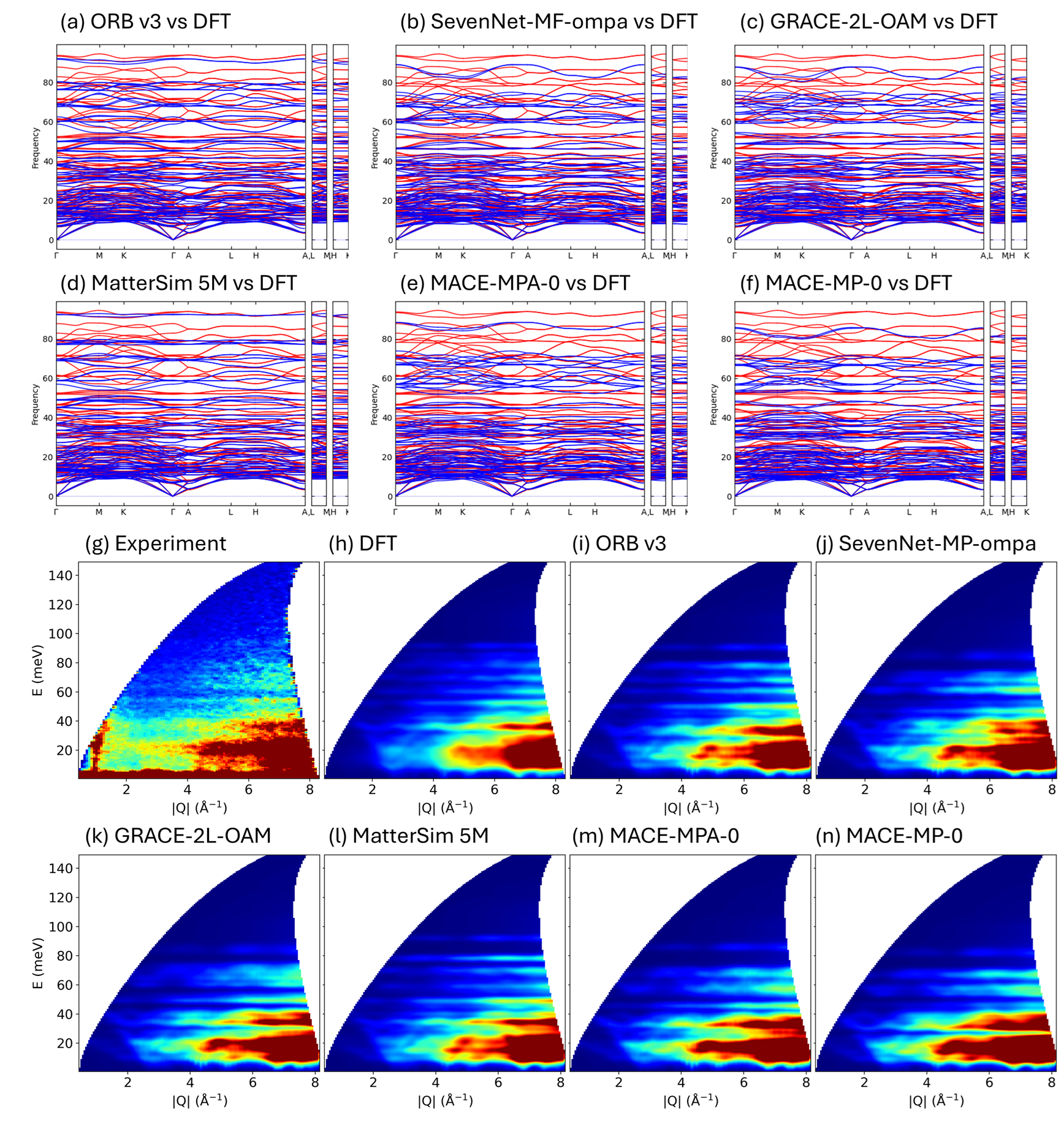}
    \caption{(a)-(f) Phonon dispersion of \ce{Ba3ZnRu2O9} obtained from uMLIPs and DFT. The powder INS spectra measured at SEQUOIA is shown in (g)\cite{cheng_direct_2023}, with the corresponding simulations in (h)-(n).}
    \label{fig:Ba3ZnRu2O9}
\end{figure}

A key advantage of INS is its unique capability in accurately measuring the complete phonon dispersion in the 4D $(\bm{Q},E)$ space. Data collected on a single crystal \ce{Cu2O} (cuprite)\cite{saunders_thermal_2022} are presented in Figure 4. Such measurements provide a detailed comparison of theoretically predicted and experimentally measured phonon dispersion along any direction of interest in the Brillouin zone. This can be considered the golden standard in the benchmarking of phonon calculations. By overlapping the calculated phonon dispersion on the experimentally measured spectra, we can clearly see that, for cuprite, MatterSim outperforms all other uMLIPs, with ORB v3 also showing excellent agreement. Underestimation of the optical band frequencies remains an issue, especially for the MACE-MP-0 potential. Another new observation is the underestimation of the acoustic branch below 10 meV, which is also present even in the DFT result. However, MACE-MPA-0 is in better agreement with the experiment on this acoustic branch.

\begin{figure}[h!tbp]
   \centering
    \includegraphics[width=0.95\textwidth]{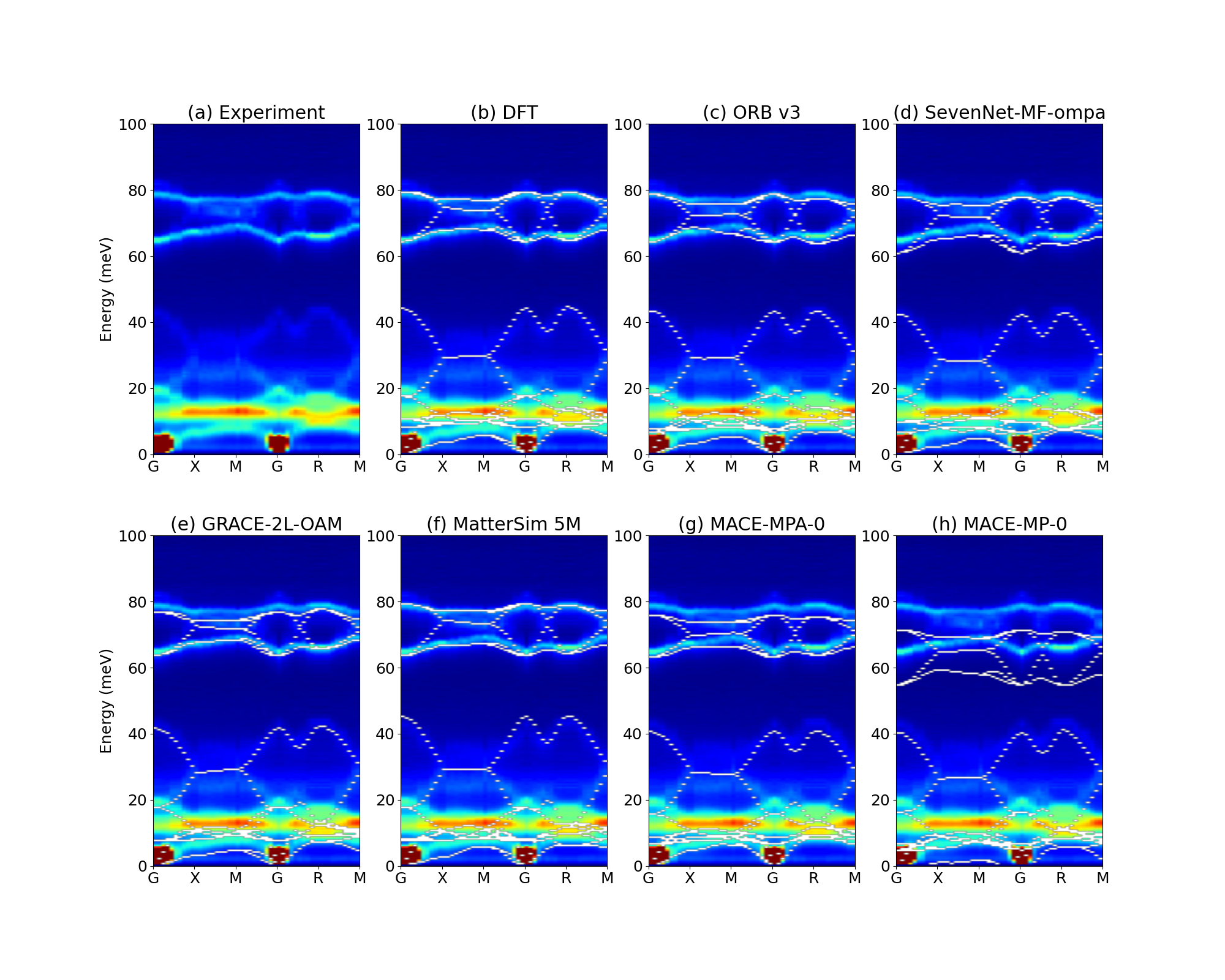}
    \caption{Phonon dispersion of \ce{Cu2O} along representative paths. (a) shows the INS spectrum obtained from a single-crystal INS experiment by Saunders \textit{et al}.\cite{saunders_thermal_2022} In (b)-(h), the calculated phonon dispersion from DFT and various uMLIPs are overlapped (as white dots) on the experimental spectrum for comparison.}
    \label{fig:cu2o}
\end{figure}

The 4D $S(\bm{Q},E)$ measured on single crystals is very sensitive to the details in phonon dispersion, and one can choose to examine the data along various directions, in multiple Brillouin zones. Figure 5 provides another example of \ce{RuCl3},\cite{mu_role_2022} where we directly compare experimental and simulated INS spectra. Such a direct comparison of the spectra illustrates the accuracy in not only the calculated phonon frequencies but also the vibrational modes. This is because the INS scattering ``intensity” is determined by the phonon polarization vectors, thus matching the simulated and experimental INS intensity profile requires not only correct frequencies, but also correct polarization vectors. In the case of \ce{RuCl3}, Figure 5 shows that MatterSim achieves near-DFT accuracy in both vibrational frequencies and modes. However, if examined closely, there are still fine details that MatterSim cannot reproduce accurately, such as the faint band at about 15 meV near the Zone center, as well as the side branches near 35 meV. Even with these discrepancies, the similarities are sufficient to allow unambiguous assignments of the modes. Other models, with varying degrees of agreement, exhibit significant discrepancies in certain parts of the spectra. 

\begin{figure}[h!tbp]
   \centering
    \includegraphics[width=0.90\textwidth]{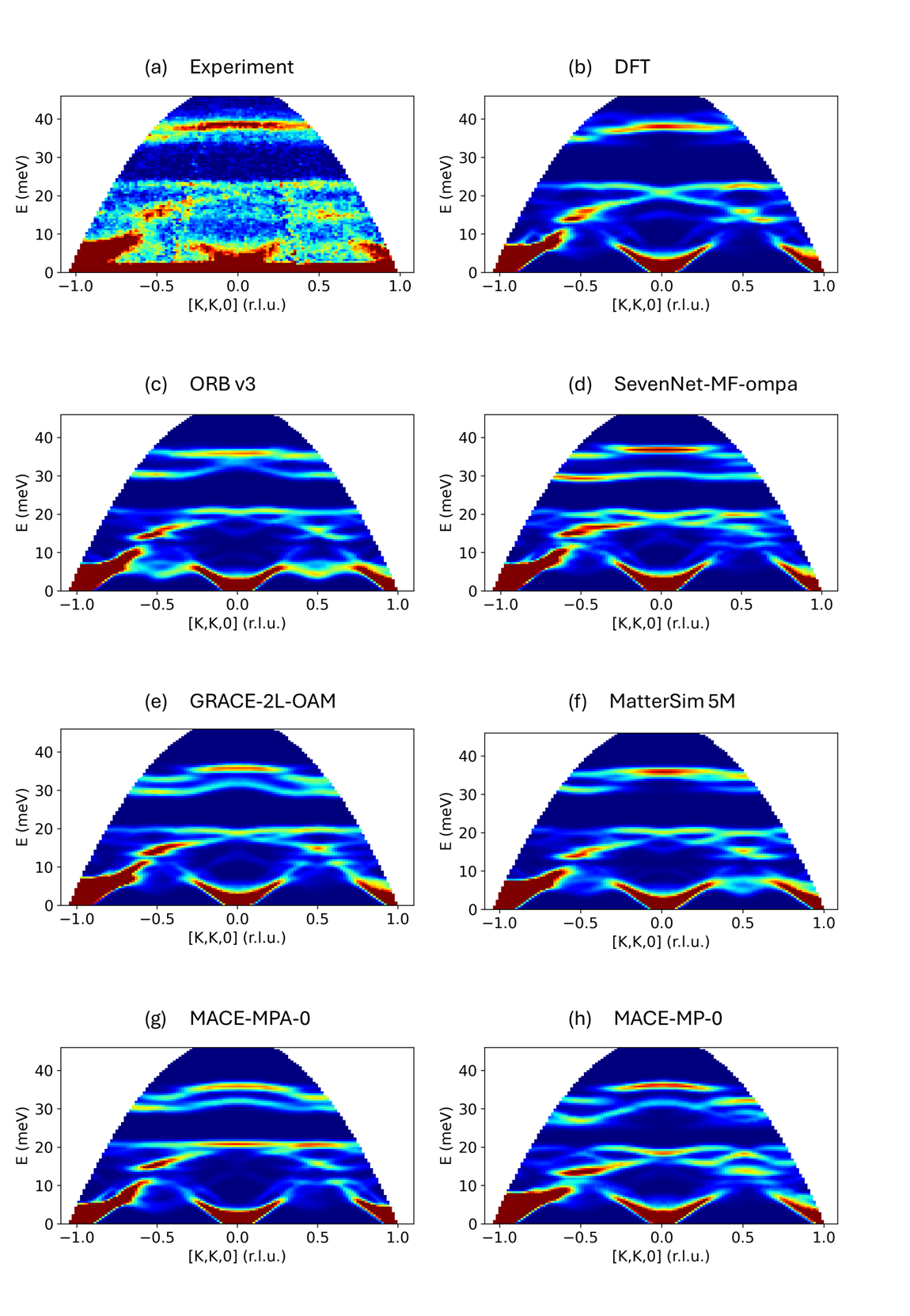}
    \caption{Single crystal INS spectra from \ce{RuCl3} along the [K,K,0] direction from (a) experiment by Mu \textit{et al}.\cite{mu_role_2022} and (b)-(h) simulations.}
    \label{fig:rucl3}
\end{figure}

We now set a higher standard to examine some more difficult cases. The modeling of hydrogen-containing materials has posed significant challenges for uMLIPs because hydrogen can form various bonds with different elements in different compounds, exhibiting drastically different vibrational properties. This is particularly relevant for INS analysis because neutrons scatter strongly from hydrogen with a large incoherent cross-section. Since the neutron scattering from hydrogen-containing material is predominantly incoherent, the $\bm{Q}$ dependence is much less relevant, and one can focus on the energy spectra in 1D VISION\cite{seeger_resolution_2009} is an indirect geometry neutron spectrometer optimized to measure such energy spectra from hydrogen-containing materials. It can be considered a neutron scattering analogue of Raman or infrared spectroscopy. In the following, we will show a few examples in Figure 6, each representing a group of hydrogen-containing materials. Specifically, we will study (a) \ce{ZrH2}, an inorganic hydride with a simple crystal structure, (b) ZIF-8, a metal-organic framework (MOF), (c) polyethylene, an organic polymer, (d) toluene, an aromatic molecular crystal, (e) butyric acid, an aliphatic hydrocarbon, and (f) remdesivir, a more complex crystal of drug molecules. In addition to the previously mentioned uMLIPs, we also include MACE-OFF (\texttt{MACE-OFF23\_medium})\cite{kovacs_mace-off_2025} to replace MACE-MP-0 when applicable. MACE-OFF is a derivative of MACE fine-tuned for organic crystals made of one or more elements in H, C, N, O, F, P, S, Cl, Br, and I.

\begin{figure}[h!tbp]
   \centering
    \includegraphics[width=0.80\textwidth]{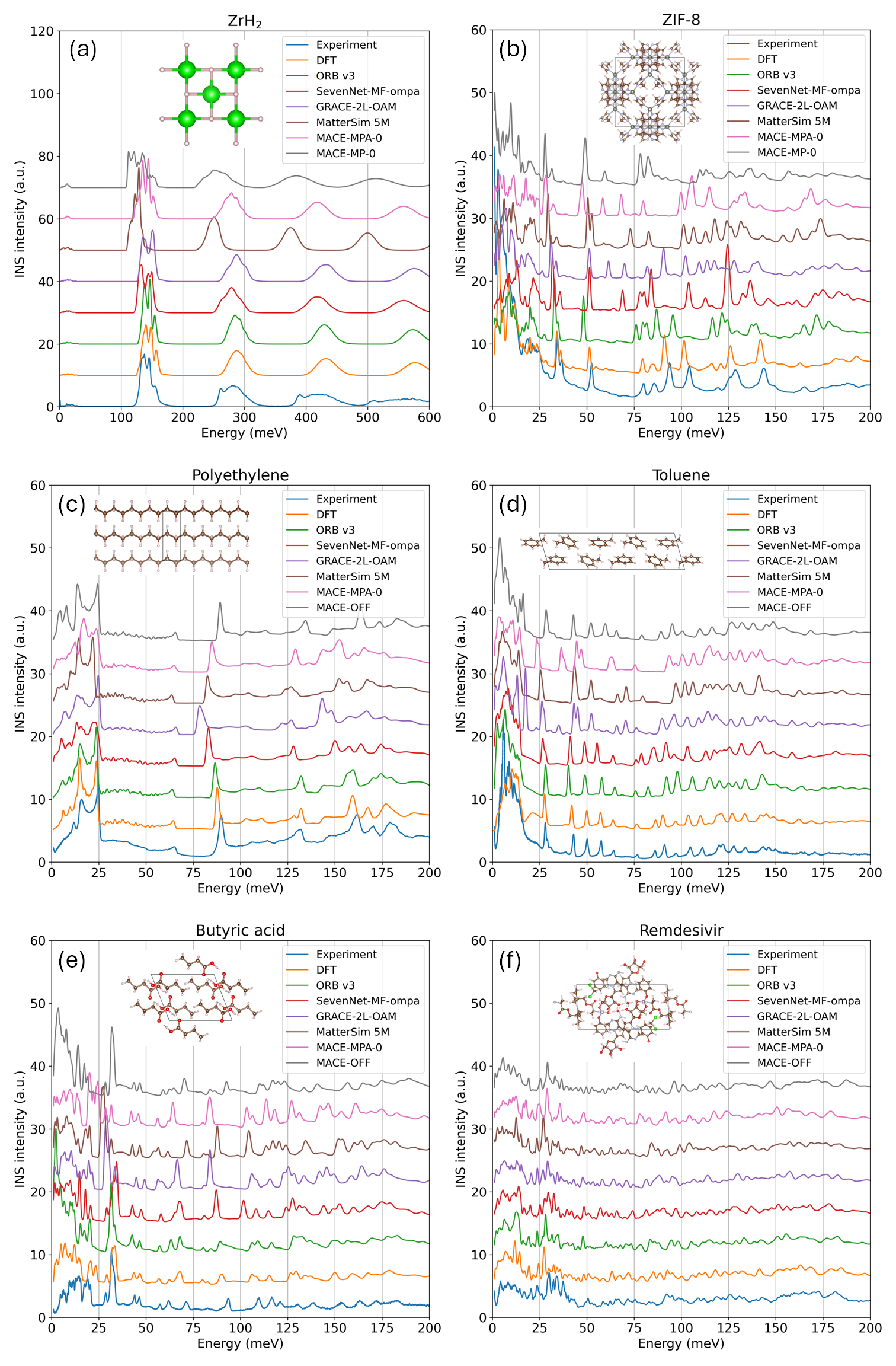}
    \caption{Experimental (measured at VISION) and simulated INS spectra for (a) \ce{ZrH2}\cite{zhang_study_2021} (b) ZIF-8\cite{casco_understanding_2017} (c) polyethylene (d) toluene\cite{cheng_direct_2023} (e) butyric acid (f) remdesivir\cite{mamontov_low_2021} Insets show the crystal structure of each material.}
    \label{fig:vision_spec}
\end{figure}

The VISION spectrum from \ce{ZrH2} (Figure 6a) is relatively simple, showing a strong \chemfig{Zr-[:0,0.6]H} libration band at 150 meV and its higher-order overtones. Although all uMLIPs reproduced this general feature, MatterSim and MACE-MP-0 underestimate frequencies significantly. The fundamental excitation peaks from MACE-MP-0 appear to be broader than they should be, possibly due to an overestimated anisotropy in the potential energy profile for hydrogen, causing broader splitting of H libration in different directions. ZIF-8 is a hybrid material that contains organic ligands anchored by metal centers. The hybrid nature makes it challenging for uMLIPs to describe. Indeed, none of the potentials tested is good enough for ZIF-8 (Figure 6b), and ORB v3 is the only one that is reasonably close. Polyethylene is a polymer with a simple composition and crystal structure. The INS peaks are mainly caused by the deformation of the molecular chain and local vibrations of the bonds in methylene groups (\ce{-CH2-}). The relative motions between chains and long-range phonon modes contribute to the band below 30 meV (Figure 6c). Not surprisingly, MACE-OFF produces the best results because it has been specifically fine-tuned to describe such interactions. ORB v3 also gives satisfactory results. Toluene has a simple molecular structure, but a relatively complex crystal structure (Figure 6d). The methyl group and the aromatic ring each produce some characteristic sharp peaks in the INS spectra, which are determined by multiple types of bonds and interactions. Figure 6d shows that only the spectrum simulated by MACE-OFF is close enough to the experiment for peak assignment and spectral interpretation. ORB v3 and SevenNet-MF-ompa also capture the lower energy features (\textless75 meV) well. The observation is similar for butyric acid (Figure 6e), where vibrations from the OH group and the short \chemfig{C-[:0,0.6]C} chain dominate the INS intensities. To further illustrate the usefulness of uMLIPs in more complex molecular crystals, we also examined remdesivir, an antiviral medication used to treat COVID-19. The spectra are much more complex with more peaks (Figure 6f), making peak assignment and spectral interpretation more challenging. The spectrum from MACE-OFF is reasonably close to that of DFT, but this is not sufficient because here even DFT fails to reproduce all the peaks.

Overall, the above analysis shows that the best performing uMLIPs can be an accurate alternative to DFT when the latter is not available or cannot meet the requirements of time-sensitive tasks. There are still certain areas where these potentials tend to be less accurate, such as inorganic-organic hybrid materials or complex molecular crystals, but they are ready to play an increasingly important role in the analysis of phonons, thermodynamic properties, and real-time spectral analysis. Specifically, we have found MatterSim 5M and ORB v3 are constant top performers in reproducing features in INS spectra. For organic materials, MACE-OFF usually performs better than others because it has been fine-tuned with databases including organic materials. The general observation is that uMLIPs trained on all three major databases (OMat24, MPtrj, and Alexandria) tend to perform better, and ORB v3 stands out in many tests, probably because it is the only one that included the complete Alexandria database with more than 30 million structures (with a total of 133 million structures). MatterSim was trained on its own database of 17 million structures. Its outstanding performance is a demonstration of the efficiency of its data selection. The coverage of a wide temperature and pressure range in their database is likely the key. Our benchmark with INS spectra also highlights that similar performance in traditional metrics (errors in energies and forces) does necessarily translate to similar performance in a specific application. In this case, we clearly show that, although high-ranking uMLIPs generally perform better in simulating INS, significant variations exist from case to case, and the best uMLIP for INS simulation may not have the highest benchmarking score with traditional metrics.

Based on the findings in this work, we have included several of the top performing uMLIPs in the new version of INSPIRED\cite{han_inspired_2024}, which has a graph user interface for users to perform fast and efficient simulations of the INS spectra. Compared to conventional DFT calculations, which take hours to days on a computer cluster, the same simulations with uMLIPs may only take seconds to minutes. Given the typical timescale of a neutron scattering experiment, the analysis and interpretation can now be performed in real-time, on-the-fly, as the experimental data are being collected, and the results can be used for immediate peak assignment, spectral interpretation, and steering of the experiment. This can have significant implications on the future of INS experiments and vibrational spectroscopy in general.

\section{Conclusion}

We have produced a high-quality DFT-based phonon database focusing on crystals with small unit cells and rich phonon dispersion features. By benchmarking uMLIPs against this database, we have demonstrated that the most recent uMLIPs, such as MatterSim, ORB v3, MACE-OFF, SevenNet-MF-ompa, GRACE-2L-OAM, and MACE-MPA-0, can potentially be used to perform calculations of phonon dispersion and vibrational spectra with near DFT accuracy. We further illustrate how the simulated INS spectra from the uMLIPs compare with experiments in various scenarios, including direct and indirect geometry spectrometers, powder and single crystal measurements, coherent and incoherent scatterers, inorganic and organic materials. The side-by-side comparisons verify the applicability of the uMLIPs for real-time analysis of INS spectra and lay the foundation for future autonomous experiments. 

The rapid development of uMLIPs has been fueled by the growth of computing power, the availability of new databases, and accelerated machine learning algorithms. It took only a few years for the early models to evolve into powerful ones that seem to be sufficient for phonon calculations. The next challenge would be the higher-order force constants, phonon anharmonicity, and properties away from equilibrium. Such properties are even more difficult to model with DFT, and the production of large-scale ground truth datasets is the key. Nonetheless, the lesson we learned from the development of uMLIPs is that AI in science is opening a new era of materials research and revolutionizing how experiments and data analysis are performed at an unprecedented speed.

\section{Methods}
\label{sec:method}

Plane-wave DFT calculations were performed using the Vienna Ab initio Simulation Package (VASP)\cite{vasp_prb_1996,paw_prb_1994}. The crystal structure from the MP database\cite{jain_commentary_2013} was taken as the initial structure, and the atomic coordinates are relaxed before phonon calculation. Perdew-Burke-Ernzerhof (PBE) implementation of the Generalized Gradient Approximation (GGA)\cite{pbe_prl_1996} was used for all calculations. The cutoff energy is 1.6 times the maximum ENMAX in the POTCAR. The density of k-points per volume is 160 \AA$^{-3}$. VASP precision level was set as ``accurate”. The energy tolerance for electronic minimization was $10^{-8}$ eV. The maximum force after relaxation was 0.002 eV/\AA. For nonmetals with more than one element, Born effective charges were calculated, and non-analytical calculations were automatically performed to describe the LO-TO splitting. Spin-polarized calculations were performed for magnetic materials. The dimensions of the supercells were automatically determined so that the minimum dimension in a supercell is at least 12 \AA. Phonopy\cite{phonopy-phono3py-JPCM, phonopy-phono3py-JPSJ} was used to calculate the dispersion and uniform sampling of phonons in the Brillouin zone. Finally, crystals with significant phonon instability after such calculations (phonons with negative/imaginary frequencies) were not included in the database. 

The specific versions of used uMLIPs are tabulated in Table S7. To calculate the phonon properties of the materials with selected uMLIPs in the database, the geometry relaxations were performed with the Atomic Simulation Environment (ASE)\cite{ase-paper}, using the fast inertial relaxation engine (FIRE) with force convergence criteria set to 0.005 ($\mathrm{eV/}$\AA). We did not relax the volume or shape of the unit cells in this study because our goal was to benchmark and compare the phonon properties calculated on the same crystal structures (unit cells). We do not evaluate the capability of the potential to predict the volume/density of the material, which corresponds to a different dimension in the potential energy profile. A proper prediction of volume/density (to be compared with experiments) also requires consideration of the temperature-dependent vibrational entropy and is deferred to future studies. 

The average difference of interatomic distance between the initial and optimized atomic coordinates for each material is calculated as
\begin{equation}
    \Delta d=\frac{\displaystyle\sum_{i=1}^{n} \left| \bm{x}_i-\hat{\bm{x}}_i \right| }{n},
\end{equation}
where $\bm{x}_i$ is the optimized atomic coordinate of atom $i$ based on DFT, $\hat{\bm{x}}_i$ is that based on uMLIP, and $n$ is the total number of atoms in the supercell of the material. Furthermore, force calculations were conducted with the ASE and Phononpy packages using finite displacement methods. The displacement is set to 0.03 \AA. 

The \textbf{q}-points for each material are generated by the pymatgen package\cite{ong_python_2013}, and the density of the grid is 1000 per \AA$^{-3}$ in the reciprocal unit cell with $\Gamma$-point at the center. The phonon frequencies are calculated on each \textbf{q}-point for all uMLIPs and DFT. The mean absolute errors (MAEs) of the phonon frequencies of each material and each uMLIP compared to that of DFT are calculated as 
\begin{equation}
    \mathrm{MAE_{\omega}}=\frac{\sum w_j\cdot\left(\frac{\sum|\omega_i-\hat{\omega}_i|}{n}\right)}{\sum w_j},
\end{equation}
where $w_j$ is the weight of \textbf{q}-point $j$, $n$ is the number of phonon modes at each \textbf{q}-point, $\omega_i$ is the frequency of phonon mode $i$ from a uMLIP and $\hat{\omega}_i$ is the frequency of phonon mode $i$ from the DFT. The phonon density of states and the thermodynamical properties including Helmholtz free energy, heat capacity, and entropy are calculated based on the phonon properties using Phonopy. For the thermodynamic properties, a temperature scan from 0 K to 1000 K is performed (every 10 K). The MAEs of the thermodynamical properties are calculated as
\begin{equation}
    \mathrm{MAE}_{X}=\frac{1}{T_{\mathrm{max}}}\int_{0}^{T_{\mathrm{max}}}|X(T)-\hat{X}(T)|\mathrm\,{\mathrm{d}}T,
\end{equation}
where the $X$ and $\hat{X}$ represent one of the thermodynamic properties from DFT and uMLIPs, respectively, $T$ is the temperature and $T_{\mathrm{max}}$ is the maximum temperature, 1000 K.

To compare the PDOS obtained from DFT calculations with those generated from uMLIP, we use Spearman’s rank correlation coefficient as the statistical measurement. Spearman’s rank correlation coefficient is expressed as
\begin{equation}
    \rho=1-\frac{6\sum_id_i^2}{n(n^2-1)},
\end{equation}
where $d_i$ is the difference between the ranks of $y_i$ and $\hat{y}_i$ in DFT PDOS and uMLIP PDOS, respectively, and $n$ is the number of elements in each vector.

\begin{acknowledgement}

The authors were supported by the Scientific User Facilities Division, Office of Basic Energy Sciences, U.S. Department of Energy (DOE), under Contract No. DE-AC0500OR22725 with UT Battelle, LLC. This research was partially sponsored by the Artificial Intelligence Initiative as part of the Laboratory Directed Research and Development (LDRD) program of Oak Ridge National Laboratory (ORNL). Computing resources were made available through the VirtuES and ICE-MAN projects, funded by the LDRD program and the Compute and Data Environment for Science (CADES) at ORNL, as well as resources of the National Energy Research Scientific Computing Center (NERSC), a U.S. Department of Energy Office of Science User Facility located at Lawrence Berkeley National Laboratory, operated under Contract No. DE-AC02-05CH11231 using NERSC award ERCAP0024340. INS spectra not previously published were obtained at the VISION beamline of the Spallation Neutron Source, a DOE Office of Science User Facility operated by ORNL. 

\end{acknowledgement}

\begin{suppinfo}

The Supporting Information is available. The Python scripts and phonon database used for the benchmarking in this study are openly available at the following URL/DOI: \url{https://github.com/maplewen4/phonon_uMLIP} and \url{https://zenodo.org/records/15298436}, respectively.

\end{suppinfo}

\bibliography{ref}

\providecommand{\latin}[1]{#1}
\makeatletter
\providecommand{\doi}
  {\begingroup\let\do\@makeother\dospecials
  \catcode`\{=1 \catcode`\}=2 \doi@aux}
\providecommand{\doi@aux}[1]{\endgroup\texttt{#1}}
\makeatother
\providecommand*\mcitethebibliography{\thebibliography}
\csname @ifundefined\endcsname{endmcitethebibliography}  {\let\endmcitethebibliography\endthebibliography}{}
\begin{mcitethebibliography}{44}
\providecommand*\natexlab[1]{#1}
\providecommand*\mciteSetBstSublistMode[1]{}
\providecommand*\mciteSetBstMaxWidthForm[2]{}
\providecommand*\mciteBstWouldAddEndPuncttrue
  {\def\EndOfBibitem{\unskip.}}
\providecommand*\mciteBstWouldAddEndPunctfalse
  {\let\EndOfBibitem\relax}
\providecommand*\mciteSetBstMidEndSepPunct[3]{}
\providecommand*\mciteSetBstSublistLabelBeginEnd[3]{}
\providecommand*\EndOfBibitem{}
\mciteSetBstSublistMode{f}
\mciteSetBstMaxWidthForm{subitem}{(\alph{mcitesubitemcount})}
\mciteSetBstSublistLabelBeginEnd
  {\mcitemaxwidthsubitemform\space}
  {\relax}
  {\relax}

\bibitem[Reissland(1973)]{reissland_physics_1973}
Reissland,~J.~A. \emph{The physics of phonons}; London ; New York : Wiley, 1973\relax
\mciteBstWouldAddEndPuncttrue
\mciteSetBstMidEndSepPunct{\mcitedefaultmidpunct}
{\mcitedefaultendpunct}{\mcitedefaultseppunct}\relax
\EndOfBibitem
\bibitem[Togo and Tanaka(2015)Togo, and Tanaka]{togo_first_2015}
Togo,~A.; Tanaka,~I. First principles phonon calculations in materials science. \emph{Scripta Materialia} \textbf{2015}, \emph{108}, 1--5\relax
\mciteBstWouldAddEndPuncttrue
\mciteSetBstMidEndSepPunct{\mcitedefaultmidpunct}
{\mcitedefaultendpunct}{\mcitedefaultseppunct}\relax
\EndOfBibitem
\bibitem[Fultz(2020)]{fultz_inelastic_2020}
Fultz,~B. \emph{Inelastic {Scattering}}, revision 1.0 ed.; California Institute of Technology, 2020\relax
\mciteBstWouldAddEndPuncttrue
\mciteSetBstMidEndSepPunct{\mcitedefaultmidpunct}
{\mcitedefaultendpunct}{\mcitedefaultseppunct}\relax
\EndOfBibitem
\bibitem[Unke \latin{et~al.}(2021)Unke, Chmiela, Sauceda, Gastegger, Poltavsky, Sch\"utt, Tkatchenko, and M\"uller]{unke_machine_2021}
Unke,~O.~T.; Chmiela,~S.; Sauceda,~H.~E.; Gastegger,~M.; Poltavsky,~I.; Sch\"utt,~K.~T.; Tkatchenko,~A.; M\"uller,~K.-R. Machine {Learning} {Force} {Fields}. \emph{Chem. Rev.} \textbf{2021}, \emph{121}, 10142--10186\relax
\mciteBstWouldAddEndPuncttrue
\mciteSetBstMidEndSepPunct{\mcitedefaultmidpunct}
{\mcitedefaultendpunct}{\mcitedefaultseppunct}\relax
\EndOfBibitem
\bibitem[Riebesell \latin{et~al.}(2024)Riebesell, Goodall, Benner, Chiang, Deng, Ceder, Asta, Lee, Jain, and Persson]{riebesell_matbench_2024}
Riebesell,~J.; Goodall,~R. E.~A.; Benner,~P.; Chiang,~Y.; Deng,~B.; Ceder,~G.; Asta,~M.; Lee,~A.~A.; Jain,~A.; Persson,~K.~A. Matbench {Discovery} -- {A} framework to evaluate machine learning crystal stability predictions. 2024; \url{http://arxiv.org/abs/2308.14920}\relax
\mciteBstWouldAddEndPuncttrue
\mciteSetBstMidEndSepPunct{\mcitedefaultmidpunct}
{\mcitedefaultendpunct}{\mcitedefaultseppunct}\relax
\EndOfBibitem
\bibitem[Deng \latin{et~al.}(2025)Deng, Choi, Zhong, Riebesell, Anand, Li, Jun, Persson, and Ceder]{deng_systematic_2025}
Deng,~B.; Choi,~Y.; Zhong,~P.; Riebesell,~J.; Anand,~S.; Li,~Z.; Jun,~K.; Persson,~K.~A.; Ceder,~G. Systematic softening in universal machine learning interatomic potentials. \emph{npj Comput Mater} \textbf{2025}, \emph{11}, 1--9\relax
\mciteBstWouldAddEndPuncttrue
\mciteSetBstMidEndSepPunct{\mcitedefaultmidpunct}
{\mcitedefaultendpunct}{\mcitedefaultseppunct}\relax
\EndOfBibitem
\bibitem[Loew \latin{et~al.}(2024)Loew, Sun, Wang, Botti, and Marques]{loew_universal_2024}
Loew,~A.; Sun,~D.; Wang,~H.-C.; Botti,~S.; Marques,~M. A.~L. Universal {Machine} {Learning} {Interatomic} {Potentials} are {Ready} for {Phonons}. 2024; \url{http://arxiv.org/abs/2412.16551}\relax
\mciteBstWouldAddEndPuncttrue
\mciteSetBstMidEndSepPunct{\mcitedefaultmidpunct}
{\mcitedefaultendpunct}{\mcitedefaultseppunct}\relax
\EndOfBibitem
\bibitem[Togo(2025)]{togo_atztogophonondb_2025}
Togo,~A. atztogo/phonondb. 2025; \url{https://github.com/atztogo/phonondb}\relax
\mciteBstWouldAddEndPuncttrue
\mciteSetBstMidEndSepPunct{\mcitedefaultmidpunct}
{\mcitedefaultendpunct}{\mcitedefaultseppunct}\relax
\EndOfBibitem
\bibitem[Han \latin{et~al.}(2024)Han, Savici, Li, and Cheng]{han_inspired_2024}
Han,~B.; Savici,~A.~T.; Li,~M.; Cheng,~Y. {INSPIRED}: {Inelastic} neutron scattering prediction for instantaneous results and experimental design. \emph{Computer Physics Communications} \textbf{2024}, \emph{304}, 109288\relax
\mciteBstWouldAddEndPuncttrue
\mciteSetBstMidEndSepPunct{\mcitedefaultmidpunct}
{\mcitedefaultendpunct}{\mcitedefaultseppunct}\relax
\EndOfBibitem
\bibitem[Jain \latin{et~al.}(2013)Jain, Ong, Hautier, Chen, Richards, Dacek, Cholia, Gunter, Skinner, Ceder, and Persson]{jain_commentary_2013}
Jain,~A.; Ong,~S.~P.; Hautier,~G.; Chen,~W.; Richards,~W.~D.; Dacek,~S.; Cholia,~S.; Gunter,~D.; Skinner,~D.; Ceder,~G. \latin{et~al.}  Commentary: {The} {Materials} {Project}: {A} materials genome approach to accelerating materials innovation. \emph{APL Materials} \textbf{2013}, \emph{1}, 011002\relax
\mciteBstWouldAddEndPuncttrue
\mciteSetBstMidEndSepPunct{\mcitedefaultmidpunct}
{\mcitedefaultendpunct}{\mcitedefaultseppunct}\relax
\EndOfBibitem
\bibitem[Fu \latin{et~al.}(2025)Fu, Wood, Barroso-Luque, Levine, Gao, Dzamba, and Zitnick]{fu_learning_2025}
Fu,~X.; Wood,~B.~M.; Barroso-Luque,~L.; Levine,~D.~S.; Gao,~M.; Dzamba,~M.; Zitnick,~C.~L. Learning {Smooth} and {Expressive} {Interatomic} {Potentials} for {Physical} {Property} {Prediction}. 2025; \url{http://arxiv.org/abs/2502.12147}\relax
\mciteBstWouldAddEndPuncttrue
\mciteSetBstMidEndSepPunct{\mcitedefaultmidpunct}
{\mcitedefaultendpunct}{\mcitedefaultseppunct}\relax
\EndOfBibitem
\bibitem[Rhodes \latin{et~al.}(2025)Rhodes, Vandenhaute, Šimkus, Gin, Godwin, Duignan, and Neumann]{rhodes2025orbv3atomisticsimulationscale}
Rhodes,~B.; Vandenhaute,~S.; Šimkus,~V.; Gin,~J.; Godwin,~J.; Duignan,~T.; Neumann,~M. Orb-v3: atomistic simulation at scale. 2025; \url{https://arxiv.org/abs/2504.06231}\relax
\mciteBstWouldAddEndPuncttrue
\mciteSetBstMidEndSepPunct{\mcitedefaultmidpunct}
{\mcitedefaultendpunct}{\mcitedefaultseppunct}\relax
\EndOfBibitem
\bibitem[Park \latin{et~al.}(2024)Park, Kim, Hwang, and Han]{park_scalable_2024}
Park,~Y.; Kim,~J.; Hwang,~S.; Han,~S. Scalable Parallel Algorithm for Graph Neural Network Interatomic Potentials in Molecular Dynamics Simulations. \emph{J. Chem. Theory Comput.} \textbf{2024}, \emph{20}, 4857--4868\relax
\mciteBstWouldAddEndPuncttrue
\mciteSetBstMidEndSepPunct{\mcitedefaultmidpunct}
{\mcitedefaultendpunct}{\mcitedefaultseppunct}\relax
\EndOfBibitem
\bibitem[Kim \latin{et~al.}(2024)Kim, Kim, Kim, Lee, Park, Kang, and Han]{kim_sevennet_mf_2024}
Kim,~J.; Kim,~J.; Kim,~J.; Lee,~J.; Park,~Y.; Kang,~Y.; Han,~S. Data-Efficient Multifidelity Training for High-Fidelity Machine Learning Interatomic Potentials. \emph{J. Am. Chem. Soc.} \textbf{2024}, \emph{147}, 1042--1054\relax
\mciteBstWouldAddEndPuncttrue
\mciteSetBstMidEndSepPunct{\mcitedefaultmidpunct}
{\mcitedefaultendpunct}{\mcitedefaultseppunct}\relax
\EndOfBibitem
\bibitem[Bochkarev \latin{et~al.}(2024)Bochkarev, Lysogorskiy, and Drautz]{bochkarev_graph_2024}
Bochkarev,~A.; Lysogorskiy,~Y.; Drautz,~R. Graph {Atomic} {Cluster} {Expansion} for {Semilocal} {Interactions} beyond {Equivariant} {Message} {Passing}. \emph{Phys. Rev. X} \textbf{2024}, \emph{14}, 021036\relax
\mciteBstWouldAddEndPuncttrue
\mciteSetBstMidEndSepPunct{\mcitedefaultmidpunct}
{\mcitedefaultendpunct}{\mcitedefaultseppunct}\relax
\EndOfBibitem
\bibitem[Yang \latin{et~al.}(2024)Yang, Hu, Zhou, Liu, Shi, Li, Li, Chen, Chen, Zeni, Horton, Pinsler, Fowler, ZÃ¼gner, Xie, Smith, Sun, Wang, Kong, Liu, Hao, and Lu]{yang_mattersim_2024}
Yang,~H.; Hu,~C.; Zhou,~Y.; Liu,~X.; Shi,~Y.; Li,~J.; Li,~G.; Chen,~Z.; Chen,~S.; Zeni,~C. \latin{et~al.}  {MatterSim}: {A} {Deep} {Learning} {Atomistic} {Model} {Across} {Elements}, {Temperatures} and {Pressures}. 2024; \url{http://arxiv.org/abs/2405.04967}\relax
\mciteBstWouldAddEndPuncttrue
\mciteSetBstMidEndSepPunct{\mcitedefaultmidpunct}
{\mcitedefaultendpunct}{\mcitedefaultseppunct}\relax
\EndOfBibitem
\bibitem[Batatia \latin{et~al.}(2024)Batatia, Benner, Chiang, Elena, Kov\'acs, Riebesell, Advincula, Asta, Avaylon, Baldwin, Berger, Bernstein, Bhowmik, Blau, CÄƒrare, Darby, De, Pia, Deringer, ElijoÅ¡ius, El-Machachi, Falcioni, Fako, Ferrari, Genreith-Schriever, George, Goodall, Grey, Grigorev, Han, Handley, Heenen, Hermansson, Holm, Jaafar, Hofmann, Jakob, Jung, Kapil, Kaplan, Karimitari, Kermode, Kroupa, Kullgren, Kuner, Kuryla, Liepuoniute, Margraf, MagdÄƒu, Michaelides, Moore, Naik, Niblett, Norwood, O'Neill, Ortner, Persson, Reuter, Rosen, Schaaf, Schran, Shi, Sivonxay, Stenczel, Svahn, Sutton, Swinburne, Tilly, Oord, Varga-Umbrich, Vegge, VondrÃ¡k, Wang, Witt, Zills, and CsÃ¡nyi]{batatia_foundation_2024}
Batatia,~I.; Benner,~P.; Chiang,~Y.; Elena,~A.~M.; Kov\'acs,~D.~P.; Riebesell,~J.; Advincula,~X.~R.; Asta,~M.; Avaylon,~M.; Baldwin,~W.~J. \latin{et~al.}  A foundation model for atomistic materials chemistry. 2024; \url{http://arxiv.org/abs/2401.00096}\relax
\mciteBstWouldAddEndPuncttrue
\mciteSetBstMidEndSepPunct{\mcitedefaultmidpunct}
{\mcitedefaultendpunct}{\mcitedefaultseppunct}\relax
\EndOfBibitem
\bibitem[Batatia \latin{et~al.}(2022)Batatia, Kovacs, Simm, Ortner, and Csanyi]{Batatia2022mace}
Batatia,~I.; Kovacs,~D.~P.; Simm,~G. N.~C.; Ortner,~C.; Csanyi,~G. {MACE}: Higher Order Equivariant Message Passing Neural Networks for Fast and Accurate Force Fields. Advances in Neural Information Processing Systems. 2022\relax
\mciteBstWouldAddEndPuncttrue
\mciteSetBstMidEndSepPunct{\mcitedefaultmidpunct}
{\mcitedefaultendpunct}{\mcitedefaultseppunct}\relax
\EndOfBibitem
\bibitem[Batatia \latin{et~al.}(2022)Batatia, Batzner, Kov{\'a}cs, Musaelian, Simm, Drautz, Ortner, Kozinsky, and Cs{\'a}nyi]{Batatia2022Design}
Batatia,~I.; Batzner,~S.; Kov{\'a}cs,~D.~P.; Musaelian,~A.; Simm,~G. N.~C.; Drautz,~R.; Ortner,~C.; Kozinsky,~B.; Cs{\'a}nyi,~G. The Design Space of E(3)-Equivariant Atom-Centered Interatomic Potentials. 2022\relax
\mciteBstWouldAddEndPuncttrue
\mciteSetBstMidEndSepPunct{\mcitedefaultmidpunct}
{\mcitedefaultendpunct}{\mcitedefaultseppunct}\relax
\EndOfBibitem
\bibitem[Barroso-Luque \latin{et~al.}(2024)Barroso-Luque, Shuaibi, Fu, Wood, Dzamba, Gao, Rizvi, Zitnick, and Ulissi]{barroso-luque_open_2024}
Barroso-Luque,~L.; Shuaibi,~M.; Fu,~X.; Wood,~B.~M.; Dzamba,~M.; Gao,~M.; Rizvi,~A.; Zitnick,~C.~L.; Ulissi,~Z.~W. Open {Materials} 2024 ({OMat24}) {Inorganic} {Materials} {Dataset} and {Models}. 2024; \url{http://arxiv.org/abs/2410.12771}\relax
\mciteBstWouldAddEndPuncttrue
\mciteSetBstMidEndSepPunct{\mcitedefaultmidpunct}
{\mcitedefaultendpunct}{\mcitedefaultseppunct}\relax
\EndOfBibitem
\bibitem[Neumann \latin{et~al.}(2024)Neumann, Gin, Rhodes, Bennett, Li, Choubisa, Hussey, and Godwin]{neumann2024orbfastscalableneural}
Neumann,~M.; Gin,~J.; Rhodes,~B.; Bennett,~S.; Li,~Z.; Choubisa,~H.; Hussey,~A.; Godwin,~J. Orb: A Fast, Scalable Neural Network Potential. 2024; \url{https://arxiv.org/abs/2410.22570}\relax
\mciteBstWouldAddEndPuncttrue
\mciteSetBstMidEndSepPunct{\mcitedefaultmidpunct}
{\mcitedefaultendpunct}{\mcitedefaultseppunct}\relax
\EndOfBibitem
\bibitem[Deng \latin{et~al.}(2023)Deng, Zhong, Jun, Riebesell, Han, Bartel, and Ceder]{deng_chgnet_2023}
Deng,~B.; Zhong,~P.; Jun,~K.; Riebesell,~J.; Han,~K.; Bartel,~C.~J.; Ceder,~G. {CHGNet} as a pretrained universal neural network potential for charge-informed atomistic modelling. \emph{Nat Mach Intell} \textbf{2023}, \emph{5}, 1031--1041\relax
\mciteBstWouldAddEndPuncttrue
\mciteSetBstMidEndSepPunct{\mcitedefaultmidpunct}
{\mcitedefaultendpunct}{\mcitedefaultseppunct}\relax
\EndOfBibitem
\bibitem[Chen and Ong(2022)Chen, and Ong]{chen_universal_2022}
Chen,~C.; Ong,~S.~P. A universal graph deep learning interatomic potential for the periodic table. \emph{Nat Comput Sci} \textbf{2022}, \emph{2}, 718--728\relax
\mciteBstWouldAddEndPuncttrue
\mciteSetBstMidEndSepPunct{\mcitedefaultmidpunct}
{\mcitedefaultendpunct}{\mcitedefaultseppunct}\relax
\EndOfBibitem
\bibitem[Baumann and Clerc(1997)Baumann, and Clerc]{baumann_computer-assisted_1997}
Baumann,~K.; Clerc,~J. Computer-assisted {IR} spectra prediction -- linked similarity searches for structures and spectra. \emph{Analytica Chimica Acta} \textbf{1997}, \emph{348}, 327--343\relax
\mciteBstWouldAddEndPuncttrue
\mciteSetBstMidEndSepPunct{\mcitedefaultmidpunct}
{\mcitedefaultendpunct}{\mcitedefaultseppunct}\relax
\EndOfBibitem
\bibitem[Zou \latin{et~al.}(2023)Zou, Zhang, Liang, Wei, Leng, Jiang, Luo, and Hu]{zou_deep_2023}
Zou,~Z.; Zhang,~Y.; Liang,~L.; Wei,~M.; Leng,~J.; Jiang,~J.; Luo,~Y.; Hu,~W. A deep learning model for predicting selected organic molecular spectra. \emph{Nat Comput Sci} \textbf{2023}, \emph{3}, 957--964\relax
\mciteBstWouldAddEndPuncttrue
\mciteSetBstMidEndSepPunct{\mcitedefaultmidpunct}
{\mcitedefaultendpunct}{\mcitedefaultseppunct}\relax
\EndOfBibitem
\bibitem[Squires(2012)]{squires_introduction_2012}
Squires,~G.~L. \emph{Introduction to the {Theory} of {Thermal} {Neutron} {Scattering}}, 3rd ed.; Cambridge University Press: Cambridge, 2012\relax
\mciteBstWouldAddEndPuncttrue
\mciteSetBstMidEndSepPunct{\mcitedefaultmidpunct}
{\mcitedefaultendpunct}{\mcitedefaultseppunct}\relax
\EndOfBibitem
\bibitem[Granroth \latin{et~al.}(2010)Granroth, Kolesnikov, Sherline, Clancy, Ross, Ruff, Gaulin, and Nagler]{granroth_sequoia_2010}
Granroth,~G.~E.; Kolesnikov,~A.~I.; Sherline,~T.~E.; Clancy,~J.~P.; Ross,~K.~A.; Ruff,~J. P.~C.; Gaulin,~B.~D.; Nagler,~S.~E. {SEQUOIA}: {A} {Newly} {Operating} {Chopper} {Spectrometer} at the {SNS}. \emph{J. Phys.: Conf. Ser.} \textbf{2010}, \emph{251}, 012058\relax
\mciteBstWouldAddEndPuncttrue
\mciteSetBstMidEndSepPunct{\mcitedefaultmidpunct}
{\mcitedefaultendpunct}{\mcitedefaultseppunct}\relax
\EndOfBibitem
\bibitem[Cheng \latin{et~al.}(2019)Cheng, Daemen, Kolesnikov, and Ramirez-Cuesta]{cheng_simulation_2019}
Cheng,~Y.~Q.; Daemen,~L.~L.; Kolesnikov,~A.~I.; Ramirez-Cuesta,~A.~J. Simulation of {Inelastic} {Neutron} {Scattering} {Spectra} {Using} {OCLIMAX}. \emph{J. Chem. Theory Comput.} \textbf{2019}, \emph{15}, 1974--1982\relax
\mciteBstWouldAddEndPuncttrue
\mciteSetBstMidEndSepPunct{\mcitedefaultmidpunct}
{\mcitedefaultendpunct}{\mcitedefaultseppunct}\relax
\EndOfBibitem
\bibitem[Cheng \latin{et~al.}(2023)Cheng, Wu, Pajerowski, Stone, Savici, Li, and Ramirez-Cuesta]{cheng_direct_2023}
Cheng,~Y.; Wu,~G.; Pajerowski,~D.~M.; Stone,~M.~B.; Savici,~A.~T.; Li,~M.; Ramirez-Cuesta,~A.~J. Direct prediction of inelastic neutron scattering spectra from the crystal structure*. \emph{Mach. Learn.: Sci. Technol.} \textbf{2023}, \emph{4}, 015010\relax
\mciteBstWouldAddEndPuncttrue
\mciteSetBstMidEndSepPunct{\mcitedefaultmidpunct}
{\mcitedefaultendpunct}{\mcitedefaultseppunct}\relax
\EndOfBibitem
\bibitem[Saunders \latin{et~al.}(2022)Saunders, Kim, Hellman, Smith, Weadock, Omelchenko, Granroth, Bernal-Choban, Lohaus, Abernathy, and Fultz]{saunders_thermal_2022}
Saunders,~C.~N.; Kim,~D.~S.; Hellman,~O.; Smith,~H.~L.; Weadock,~N.~J.; Omelchenko,~S.~T.; Granroth,~G.~E.; Bernal-Choban,~C.~M.; Lohaus,~S.~H.; Abernathy,~D.~L. \latin{et~al.}  Thermal expansion and phonon anharmonicity of cuprite studied by inelastic neutron scattering and \textit{ab initio} calculations. \emph{Phys. Rev. B} \textbf{2022}, \emph{105}, 174308\relax
\mciteBstWouldAddEndPuncttrue
\mciteSetBstMidEndSepPunct{\mcitedefaultmidpunct}
{\mcitedefaultendpunct}{\mcitedefaultseppunct}\relax
\EndOfBibitem
\bibitem[Mu \latin{et~al.}(2022)Mu, Dixit, Wang, Abernathy, Cao, Nagler, Yan, Lampen-Kelley, Mandrus, Polanco, Liang, HalÃ¡sz, Cheng, Banerjee, and Berlijn]{mu_role_2022}
Mu,~S.; Dixit,~K.~D.; Wang,~X.; Abernathy,~D.~L.; Cao,~H.; Nagler,~S.~E.; Yan,~J.; Lampen-Kelley,~P.; Mandrus,~D.; Polanco,~C.~A. \latin{et~al.}  Role of the third dimension in searching for {Majorana} fermions in $\alpha$-\ce{RuCl3} via phonons. \emph{Phys. Rev. Research} \textbf{2022}, \emph{4}, 013067\relax
\mciteBstWouldAddEndPuncttrue
\mciteSetBstMidEndSepPunct{\mcitedefaultmidpunct}
{\mcitedefaultendpunct}{\mcitedefaultseppunct}\relax
\EndOfBibitem
\bibitem[Seeger \latin{et~al.}(2009)Seeger, Daemen, and Larese]{seeger_resolution_2009}
Seeger,~P.~A.; Daemen,~L.~L.; Larese,~J.~Z. Resolution of {VISION}, a crystal-analyzer spectrometer. \emph{Nuclear Instruments and Methods in Physics Research Section A: Accelerators, Spectrometers, Detectors and Associated Equipment} \textbf{2009}, \emph{604}, 719--728\relax
\mciteBstWouldAddEndPuncttrue
\mciteSetBstMidEndSepPunct{\mcitedefaultmidpunct}
{\mcitedefaultendpunct}{\mcitedefaultseppunct}\relax
\EndOfBibitem
\bibitem[Kov\'acs \latin{et~al.}(2025)Kov\'acs, Moore, Browning, Batatia, Horton, Pu, Kapil, Witt, Magd$\breve{\text{a}}$u, Cole, and Cs\'anyi]{kovacs_mace-off_2025}
Kov\'acs,~D.~P.; Moore,~J.~H.; Browning,~N.~J.; Batatia,~I.; Horton,~J.~T.; Pu,~Y.; Kapil,~V.; Witt,~W.~C.; Magd$\breve{\text{a}}$u,~I.-B.; Cole,~D.~J. \latin{et~al.}  {MACE}-{OFF}: {Transferable} {Short} {Range} {Machine} {Learning} {Force} {Fields} for {Organic} {Molecules}. 2025; \url{http://arxiv.org/abs/2312.15211}\relax
\mciteBstWouldAddEndPuncttrue
\mciteSetBstMidEndSepPunct{\mcitedefaultmidpunct}
{\mcitedefaultendpunct}{\mcitedefaultseppunct}\relax
\EndOfBibitem
\bibitem[Zhang \latin{et~al.}(2021)Zhang, Cheng, Kolesnikov, Bernholc, Lu, and Ramirez-Cuesta]{zhang_study_2021}
Zhang,~J.; Cheng,~Y.; Kolesnikov,~A.~I.; Bernholc,~J.; Lu,~W.; Ramirez-Cuesta,~A.~J. Study of {Anharmonicity} in {Zirconium} {Hydrides} {Using} {Inelastic} {Neutron} {Scattering} and {Ab}-{Initio} {Computer} {Modeling}. \emph{Inorganics} \textbf{2021}, \emph{9}, 29\relax
\mciteBstWouldAddEndPuncttrue
\mciteSetBstMidEndSepPunct{\mcitedefaultmidpunct}
{\mcitedefaultendpunct}{\mcitedefaultseppunct}\relax
\EndOfBibitem
\bibitem[Casco \latin{et~al.}(2017)Casco, Fern\'andez-Catal\'a, Cheng, Daemen, Ramirez-Cuesta, Cuadrado-Collados, Silvestre-Albero, and Ramos-Fernandez]{casco_understanding_2017}
Casco,~M.~E.; Fern\'andez-Catal\'a,~J.; Cheng,~Y.; Daemen,~L.; Ramirez-Cuesta,~A.~J.; Cuadrado-Collados,~C.; Silvestre-Albero,~J.; Ramos-Fernandez,~E.~V. Understanding {ZIF}-8 {Performance} upon {Gas} {Adsorption} by {Means} of {Inelastic} {Neutron} {Scattering}. \emph{ChemistrySelect} \textbf{2017}, \emph{2}, 2750--2753\relax
\mciteBstWouldAddEndPuncttrue
\mciteSetBstMidEndSepPunct{\mcitedefaultmidpunct}
{\mcitedefaultendpunct}{\mcitedefaultseppunct}\relax
\EndOfBibitem
\bibitem[Mamontov \latin{et~al.}(2021)Mamontov, Cheng, Daemen, Kolesnikov, Ramirez-Cuesta, Ryder, and Stone]{mamontov_low_2021}
Mamontov,~E.; Cheng,~Y.; Daemen,~L.~L.; Kolesnikov,~A.~I.; Ramirez-Cuesta,~A.~J.; Ryder,~M.~R.; Stone,~M.~B. Low rotational barriers for the most dynamically active methyl groups in the proposed antiviral drugs for treatment of {SARS}-{CoV}-2, apilimod and tetrandrine. \emph{Chemical Physics Letters} \textbf{2021}, \emph{777}, 138727\relax
\mciteBstWouldAddEndPuncttrue
\mciteSetBstMidEndSepPunct{\mcitedefaultmidpunct}
{\mcitedefaultendpunct}{\mcitedefaultseppunct}\relax
\EndOfBibitem
\bibitem[Kresse and Furthm\"uller(1996)Kresse, and Furthm\"uller]{vasp_prb_1996}
Kresse,~G.; Furthm\"uller,~J. Efficient iterative schemes for ab initio total-energy calculations using a plane-wave basis set. \emph{Phys. Rev. B} \textbf{1996}, \emph{54}, 11169--11186\relax
\mciteBstWouldAddEndPuncttrue
\mciteSetBstMidEndSepPunct{\mcitedefaultmidpunct}
{\mcitedefaultendpunct}{\mcitedefaultseppunct}\relax
\EndOfBibitem
\bibitem[Bl\"ochl(1994)]{paw_prb_1994}
Bl\"ochl,~P.~E. Projector augmented-wave method. \emph{Phys. Rev. B} \textbf{1994}, \emph{50}, 17953--17979\relax
\mciteBstWouldAddEndPuncttrue
\mciteSetBstMidEndSepPunct{\mcitedefaultmidpunct}
{\mcitedefaultendpunct}{\mcitedefaultseppunct}\relax
\EndOfBibitem
\bibitem[Perdew \latin{et~al.}(1996)Perdew, Burke, and Ernzerhof]{pbe_prl_1996}
Perdew,~J.~P.; Burke,~K.; Ernzerhof,~M. Generalized Gradient Approximation Made Simple. \emph{Phys. Rev. Lett.} \textbf{1996}, \emph{77}, 3865--3868\relax
\mciteBstWouldAddEndPuncttrue
\mciteSetBstMidEndSepPunct{\mcitedefaultmidpunct}
{\mcitedefaultendpunct}{\mcitedefaultseppunct}\relax
\EndOfBibitem
\bibitem[Togo \latin{et~al.}(2023)Togo, Chaput, Tadano, and Tanaka]{phonopy-phono3py-JPCM}
Togo,~A.; Chaput,~L.; Tadano,~T.; Tanaka,~I. Implementation strategies in phonopy and phono3py. \emph{J. Phys. Condens. Matter} \textbf{2023}, \emph{35}, 353001\relax
\mciteBstWouldAddEndPuncttrue
\mciteSetBstMidEndSepPunct{\mcitedefaultmidpunct}
{\mcitedefaultendpunct}{\mcitedefaultseppunct}\relax
\EndOfBibitem
\bibitem[Togo(2023)]{phonopy-phono3py-JPSJ}
Togo,~A. First-principles Phonon Calculations with Phonopy and Phono3py. \emph{J. Phys. Soc. Jpn.} \textbf{2023}, \emph{92}, 012001\relax
\mciteBstWouldAddEndPuncttrue
\mciteSetBstMidEndSepPunct{\mcitedefaultmidpunct}
{\mcitedefaultendpunct}{\mcitedefaultseppunct}\relax
\EndOfBibitem
\bibitem[Larsen \latin{et~al.}(2017)Larsen, Mortensen, Blomqvist, Castelli, Christensen, Dułak, Friis, Groves, Hammer, Hargus, Hermes, Jennings, Jensen, Kermode, Kitchin, Kolsbjerg, Kubal, Kaasbjerg, Lysgaard, Maronsson, Maxson, Olsen, Pastewka, Peterson, Rostgaard, Schiøtz, Schütt, Strange, Thygesen, Vegge, Vilhelmsen, Walter, Zeng, and Jacobsen]{ase-paper}
Larsen,~A.~H.; Mortensen,~J.~J.; Blomqvist,~J.; Castelli,~I.~E.; Christensen,~R.; Dułak,~M.; Friis,~J.; Groves,~M.~N.; Hammer,~B.; Hargus,~C. \latin{et~al.}  The atomic simulation environment—a Python library for working with atoms. \emph{Journal of Physics: Condensed Matter} \textbf{2017}, \emph{29}, 273002\relax
\mciteBstWouldAddEndPuncttrue
\mciteSetBstMidEndSepPunct{\mcitedefaultmidpunct}
{\mcitedefaultendpunct}{\mcitedefaultseppunct}\relax
\EndOfBibitem
\bibitem[Ong \latin{et~al.}(2013)Ong, Richards, Jain, Hautier, Kocher, Cholia, Gunter, Chevrier, Persson, and Ceder]{ong_python_2013}
Ong,~S.~P.; Richards,~W.~D.; Jain,~A.; Hautier,~G.; Kocher,~M.; Cholia,~S.; Gunter,~D.; Chevrier,~V.~L.; Persson,~K.~A.; Ceder,~G. Python {Materials} {Genomics} (pymatgen): {A} robust, open-source python library for materials analysis. \emph{Computational Materials Science} \textbf{2013}, \emph{68}, 314--319\relax
\mciteBstWouldAddEndPuncttrue
\mciteSetBstMidEndSepPunct{\mcitedefaultmidpunct}
{\mcitedefaultendpunct}{\mcitedefaultseppunct}\relax
\EndOfBibitem
\end{mcitethebibliography}

\end{document}


\vfill
This manuscript has been authored by UT-Battelle, LLC under Contract No. DE-AC05-00OR22725 with the U.S. Department of Energy.  The United States Government retains and the publisher, by accepting the article for publication, acknowledges that the United States Government retains a non-exclusive, paid-up, irrevocable, world-wide license to publish or reproduce the published form of this manuscript, or allow others to do so, for United States Government purposes.  The Department of Energy will provide public access to these results of federally sponsored research in accordance with the DOE Public Access Plan (http://energy.gov/downloads/doe-public-access-plan).

\clearpage




\begin{figure}[h!tbp]
   \centering
    \includegraphics[width=1.00\textwidth]{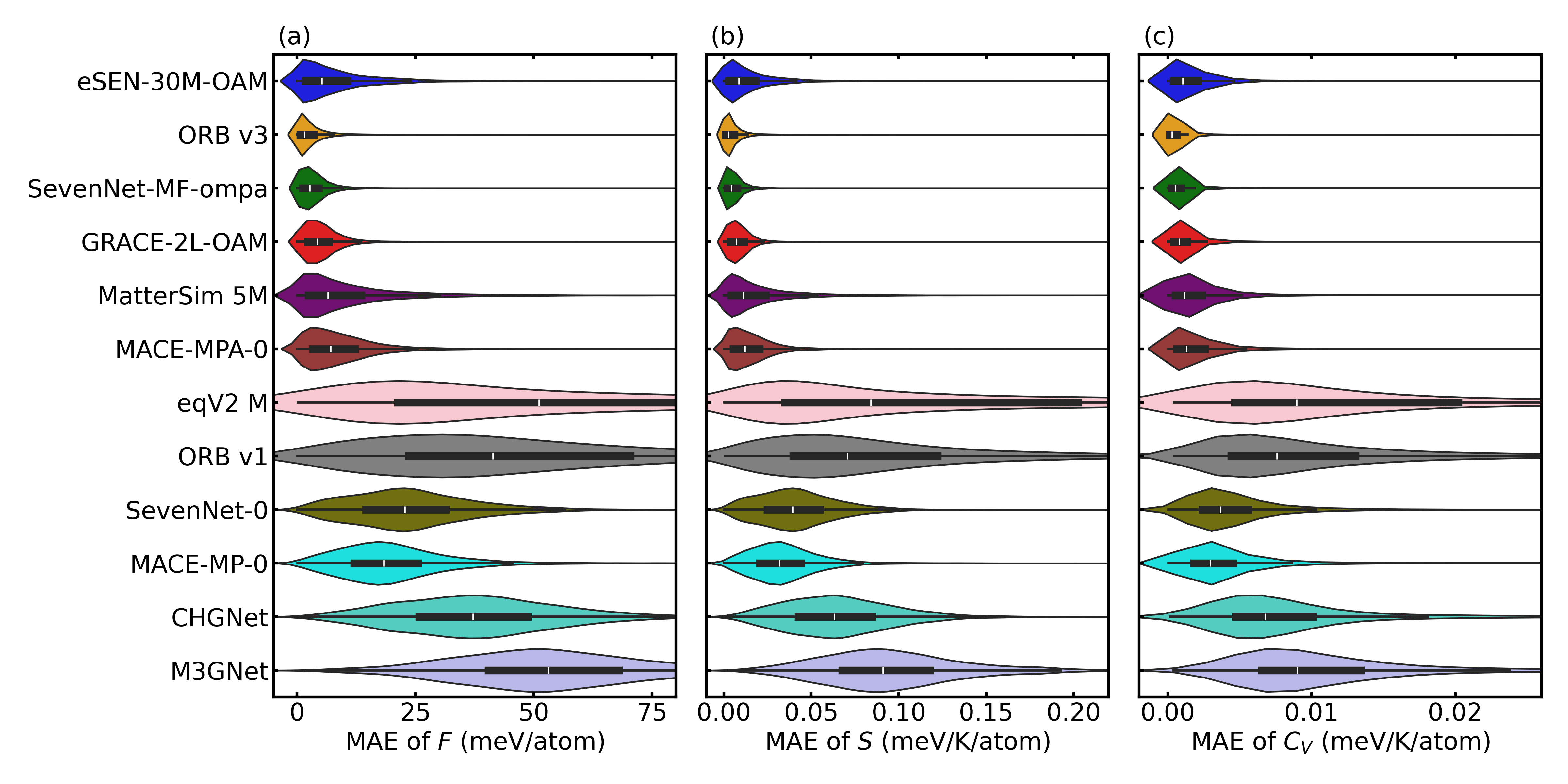}
    \caption{Violin plots of the mean absolute errors (MAEs) of (a) Helmholtz free energy ($F$), (b) entropy ($S$), and (c) constant volume heat capacity ($C_V$) for the 12 uMLIPs benchmarked against DFT.}
    \label{fig:violin}
\end{figure}


\begin{table}
\center
\caption{Statistics (over the ~5000 benchmark DFT calculations) of the differences in optimized atomic coordinates, in the units of \AA. STD stands for standard deviation.}
\begin{tabular}{cccccc}
\toprule
\textbf{uMLIP} & \textbf{Mean} & \textbf{Median} & \textbf{Min} & \textbf{Max} & \textbf{STD} \\
\midrule
eSEN-30M-OAM & 0.001697 & 0.000359 & 0.000000 & 0.057071 & 0.003548 \\
ORB v3 & 0.003686 & 0.000980 & 0.000000 & 0.116093 & 0.007262 \\
SevenNet-MF-ompa & 0.003657 & 0.001529 & 0.000000 & 0.094355 & 0.006326 \\
GRACE-2L-OAM & 0.002655 & 0.001093 & 0.000000 & 0.064231 & 0.004825 \\
MatterSim 5M & 0.006776 & 0.002302 & 0.000000 & 0.580776 & 0.022039 \\
MACE-MPA-0 & 0.004439 & 0.001809 & 0.000000 & 0.109516 & 0.007369 \\
eqV2 M & 0.001030 & 0.000000 & 0.000000 & 0.327452 & 0.005470 \\
ORB v1 & 0.007148 & 0.002542 & 0.000000 & 0.512366 & 0.020183 \\
SevenNet-0 & 0.008831 & 0.003323 & 0.000000 & 0.578190 & 0.021554 \\
MACE-MP-0 & 0.005814 & 0.001815 & 0.000000 & 0.272337 & 0.011769 \\
CHGNet & 0.014086 & 0.005743 & 0.000000 & 0.499643 & 0.027146 \\
M3GNet & 0.016768 & 0.005493 & 0.000000 & 0.543423 & 0.034797 \\
\bottomrule
\end{tabular}
\end{table}

\begin{table}
\center
\caption{Statistics of the MAE of frequencies, in the units of meV.}
\begin{tabular}{cccccc}
\toprule
\textbf{uMLIP} & \textbf{Mean} & \textbf{Median} & \textbf{Min} & \textbf{Max} & \textbf{STD} \\
\midrule
eSEN-30M-OAM & 1.197632 & 0.841443 & 0.021130 & 18.634622 & 1.241485 \\
ORB v3 & 0.502770 & 0.358956 & 0.020645 & 13.115362 & 0.571953 \\
SevenNet-MF-ompa & 0.710866 & 0.502641 & 0.053233 & 17.358618 & 0.763854 \\
GRACE-2L-OAM & 0.904241 & 0.663145 & 0.074595 & 11.968045 & 0.853647 \\
MatterSim 5M & 1.430077 & 1.004587 & 0.141981 & 73.417662 & 1.718469 \\
MACE-MPA-0 & 1.605055 & 1.064543 & 0.090817 & 36.284184 & 1.758067 \\
eqV2 M & 9.781785 & 6.587955 & 0.371663 & 107.770612 & 9.933934 \\
ORB v1 & 7.465269 & 5.488930 & 0.897933 & 65.137556 & 6.286905 \\
SevenNet-0 & 3.373666 & 2.670661 & 0.181043 & 47.368970 & 2.890712 \\
MACE-MP-0 & 2.899416 & 2.157099 & 0.140433 & 61.173875 & 2.849692 \\
CHGNet & 5.486877 & 4.579496 & 0.252557 & 70.120811 & 4.110288 \\
M3GNet & 7.619456 & 6.298240 & 0.453717 & 91.832068 & 5.442797 \\
\bottomrule
\end{tabular}
\end{table}

\begin{table}

\centering
\caption{Statistics of the Spearman coefficients of the phonon density of states.}
\begin{tabular}{cccccc}
\toprule
\textbf{uMLIP} & \textbf{Mean} & \textbf{Median} & \textbf{Min} & \textbf{Max} & \textbf{STD} \\
\midrule
eSEN-30M-OAM & 0.912421 & 0.939999 & 0.199052 & 0.999999 & 0.088638 \\
ORB v3 & 0.956277 & 0.971095 & 0.489224 & 1.000000 & 0.048809 \\
SevenNet-MF-ompa & 0.944057 & 0.960185 & 0.460636 & 0.999999 & 0.054207 \\
GRACE-2L-OAM & 0.932523 & 0.949827 & 0.324536 & 0.999998 & 0.062328 \\
MatterSim 5M & 0.902993 & 0.924428 & 0.152063 & 0.999992 & 0.087567 \\
MACE-MPA-0 & 0.894755 & 0.919868 & 0.110088 & 0.999998 & 0.089345 \\
eqV2 M & 0.608265 & 0.653092 & -0.046604 & 0.999968 & 0.220875 \\
ORB v1 & 0.663121 & 0.691565 & -0.077602 & 0.980417 & 0.156574 \\
SevenNet-0 & 0.837309 & 0.861869 & 0.145431 & 0.999968 & 0.111206 \\
MACE-MP-0 & 0.850970 & 0.875203 & -0.003241 & 0.999988 & 0.112676 \\
CHGNet & 0.767218 & 0.787911 & 0.111250 & 0.999982 & 0.132372 \\
M3GNet & 0.723218 & 0.744222 & 0.069009 & 0.999968 & 0.134316 \\
\bottomrule
\end{tabular}
\end{table}

\begin{table}
\center
\caption{Statistics of the MAE of $F$, in the units of meV/atom.}
\begin{tabular}{cccccc}
\toprule
\textbf{uMLIP}& \textbf{Mean} & \textbf{Median} & \textbf{Min} & \textbf{Max} & \textbf{STD} \\
\midrule
eSEN-30M-OAM & 7.903143 & 5.194921 & 0.001963 & 226.282410 & 9.320084 \\
ORB v3 & 2.897484 & 1.548290 & 0.002629 & 138.385523 & 5.044951 \\
SevenNet-MF-ompa & 3.493488 & 2.639756 & 0.009196 & 194.742176 & 4.400046 \\
GRACE-2L-OAM & 5.052455 & 4.287139 & 0.012294 & 190.051100 & 4.840543 \\
MatterSim 5M & 10.314748 & 6.502003 & 0.007308 & 287.230198 & 12.555343 \\
MACE-MPA-0 & 8.981450 & 7.053769 & 0.013711 & 195.481955 & 8.797535 \\
eqV2 M & 79.327495 & 51.099265 & 0.180236 & 428.757627 & 74.938929 \\
ORB v1 & 55.093116 & 41.344929 & 0.107440 & 396.892342 & 47.558486 \\
SevenNet-0 & 24.011399 & 22.747982 & 0.022336 & 141.687235 & 13.370331 \\
MACE-MP-0 & 19.912102 & 18.315977 & 0.124850 & 255.769144 & 12.229458 \\
CHGNet & 38.206520 & 37.176686 & 0.083159 & 222.025518 & 17.628349 \\
M3GNet & 55.584096 & 53.118416 & 1.989433 & 204.821826 & 22.774852 \\
\bottomrule
\end{tabular}
\end{table}

\begin{table}
\center
\caption{Statistics of the MAE of $S$, in the units of meV/K/atom.}
\begin{tabular}{cccccc}
\toprule
\textbf{uMLIP} & \textbf{Mean} & \textbf{Median} & \textbf{Min} & \textbf{Max} & \textbf{STD} \\
\midrule
eSEN-30M-OAM & 0.013602 & 0.008578 & 0.000005 & 0.556347 & 0.017690 \\
ORB v3 & 0.005213 & 0.002672 & 0.000007 & 0.330043 & 0.010322 \\
SevenNet-MF-ompa & 0.005922 & 0.004446 & 0.000021 & 0.484419 & 0.009020 \\
GRACE-2L-OAM & 0.008548 & 0.007238 & 0.000032 & 0.487593 & 0.009677 \\
MatterSim 5M & 0.018374 & 0.011220 & 0.000024 & 0.412023 & 0.023184 \\
MACE-MPA-0 & 0.015035 & 0.011999 & 0.000035 & 0.409624 & 0.015399 \\
eqV2 M & 0.136176 & 0.084142 & 0.000466 & 0.800892 & 0.134869 \\
ORB v1 & 0.095371 & 0.070712 & 0.000687 & 0.754481 & 0.085322 \\
SevenNet-0 & 0.041626 & 0.039555 & 0.000124 & 0.349409 & 0.024020 \\
MACE-MP-0 & 0.034233 & 0.031794 & 0.000160 & 0.650549 & 0.021579 \\
CHGNet & 0.065926 & 0.063268 & 0.000417 & 0.389369 & 0.032971 \\
M3GNet & 0.095651 & 0.091073 & 0.002833 & 0.521836 & 0.041738 \\
\bottomrule
\end{tabular}
\end{table}

\begin{table}
\center
\caption{Statistics of the MAE of $C_V$, in the units of meV/K/atom.}
\begin{tabular}{cccccc}
\toprule
\textbf{uMLIP} & \textbf{Mean} & \textbf{Median} & \textbf{Min} & \textbf{Max} & \textbf{STD} \\
\midrule
eSEN-30M-OAM & 0.001590 & 0.001054 & 0.000002 & 0.191969 & 0.003707 \\
ORB v3 & 0.000654 & 0.000294 & 0.000005 & 0.101397 & 0.002867 \\
SevenNet-MF-ompa & 0.000814 & 0.000516 & 0.000006 & 0.173494 & 0.002696 \\
GRACE-2L-OAM & 0.001113 & 0.000786 & 0.000005 & 0.193098 & 0.002984 \\
MatterSim 5M & 0.002044 & 0.001157 & 0.000014 & 0.169632 & 0.005512 \\
MACE-MPA-0 & 0.002012 & 0.001306 & 0.000013 & 0.204820 & 0.003682 \\
eqV2 M & 0.016443 & 0.008954 & 0.000427 & 0.239700 & 0.019587 \\
ORB v1 & 0.010421 & 0.007591 & 0.000438 & 0.222399 & 0.010883 \\
SevenNet-0 & 0.004761 & 0.003662 & 0.000042 & 0.161872 & 0.005542 \\
MACE-MP-0 & 0.003884 & 0.002953 & 0.000054 & 0.246958 & 0.005667 \\
CHGNet & 0.009736 & 0.006770 & 0.000155 & 0.162913 & 0.010894 \\
M3GNet & 0.011969 & 0.008997 & 0.000368 & 0.202097 & 0.011245 \\
\bottomrule
\end{tabular}
\end{table}

\begin{table}
\centering
\begin{tabular}{cc}
\hline
\textbf{uMLIP} & \textbf{Specific model} \\
\hline
eSEN-30M-OAM & \texttt{esen\_30m\_oam} \\
ORB v3 & \texttt{orb\_v3\_conservative\_inf\_omat}  \\
SevenNet-MF-ompa & \texttt{7net-mf-ompa}  \\
GRACE-2L-OAM & \texttt{GRACE-2L-OAM}  \\
MatterSim 5M & \texttt{mattersim-v1.0.0-5M} \\
MACE-MPA-0 & \texttt{mace-mpa-0-medium} \\
eqV2 M & \texttt{eqV2\_153M\_omat\_mp\_salex}  \\
ORB v1 & \texttt{orbff-v1-20240827} \\
SevenNet-0 & \texttt{SevenNet-0\_11July2024}  \\
MACE-MP-0 & \texttt{2024-01-07-mace-128-L2\_epoch-199}  \\
CHGNet & \texttt{CHGNet-MPtrj-2024.2.13-PES-11M}  \\
M3GNet & \texttt{M3GNet-MP-2021.2.8-PES} \\
\hline
\end{tabular}
\caption{List of uMLIPs and their corresponding versions.}
\label{tab:model_versions}
\end{table}